\documentclass[sigconf, nonacm]{acmart}

\newcommand*\circled[1]{\textcircled{\raisebox{-0.9pt}{#1}}}
\DeclareMathOperator*{\argmax}{arg\,max}

\usepackage{algorithm}
\usepackage{algpseudocode}
\usepackage{cleveref}

\begin{document}
\title{Towards Multi-Modal DBMSs\\for Seamless Querying of Texts and Tables}

\author{Matthias Urban}
\affiliation{
  \institution{TU Darmstadt}
  \country{Germany}}
\email{matthias.urban@cs.tu-darmstadt.de}

\author{Carsten Binnig}
\affiliation{
  \institution{TU Darmstadt \& DFKI}
  \country{Germany}}
\email{carsten.binnig@cs.tu-darmstadt.de}

\begin{abstract}
In this paper, we propose Multi-Modal Databases (MMDBs), which is a new class of database systems that can seamlessly query text and tables  using SQL.
To enable seamless querying of textual data using SQL in an MMDB, we propose to extend relational databases with so-called multi-modal operators (MMOps) which are based on the advances of recent large language models such as GPT-3.
The main idea of MMOps is that they allow text collections to be treated as tables without the need to manually transform the data.
As we show in our evaluation, our MMDB prototype can not only outperform state-of-the-art approaches such as text-to-table in terms of accuracy and performance but it also requires significantly less training data to fine-tune the model for an unseen text collection.
\end{abstract}

\maketitle

\section{Introduction}

\noindent\textbf{More than Tables.} Decades of research have turned relational databases into highly optimized systems for managing tabular data.
However, modern data applications need to deal with other data modalities as well that are often used in addition to tabular data, such as texts or image data \cite{symphony,neuraldb,wannadb}.
Unfortunately, traditional relational databases are not well-equipped to handle these multi-modal scenarios.
Today, modalities other than tables are usually not well supported inside relational databases, making it hard to reason about multi-modal datasets.
At the same time, rapid advancements in natural language processing and computer vision have made it easier to extract insights from texts, images, and other modalities.

In light of these developments, we believe that it is time to bring these innovations to the world of databases to enable the querying of not only tabular but all types of other data sources.
Although some extensions have been integrated into database systems such as full-text search or pattern matching for textual data \cite{full-text-search-microsoft}, these other modalities do by far not allow for the same level of querying via SQL as tabular data.
Our work aims to fill this gap in the field.
To this end, in this paper, we propose Multi-Modal Databases (MMDBs), which is a new class of database systems that can seamlessly store and process data of other modalities as if they would be transformed to tables but without the need to transform them in a first place.

\begin{figure}[t]
 \centering
 \includegraphics[width=0.9\linewidth]{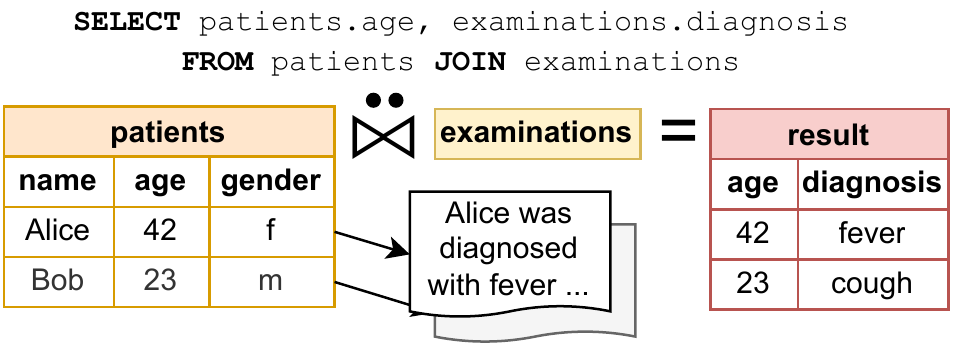}
 \vspace{-2.5ex}
 \caption{Example of a SQL Query that executes a multi-modal join between a patient table and examination reports. The multi-modal join analyzes the text and extracts values for each queried attribute such as the diagnosis from each examination report.}
 \vspace{-4.5ex}
 \label{join-example}
\end{figure}

\noindent\textbf{A Simple Example.} 
Figure \ref{join-example} illustrates how we envision how such an MMDB can be used by applications.
In the example, the MMDB stores structured patient information as a table that is linked to textual reports that contain additional diagnostic information.
If the diagnostic information would be stored in tabular form as well, a SQL query as shown at the top of Figure \ref{join-example} could easily be used to analyze the correlation between the patient's age and their diagnosis.
If the information about the diagnosis is, however, stored inside textual reports that contain the data in an unstructured form, today a data analyst would first need to create a complex data extraction pipeline to retrieve the diagnostic information from the textual reports.

\noindent\textbf{Multi-Modal Operators.} 
To enable seamless querying of multi-modal data using SQL without extracting tables in the first place, in this paper, we propose to extend relational databases with so-called multi-modal operators (MMOps).
The basic idea of MMOps is that they extend the set of operators used in traditional relational database systems by new operators that can natively ingest and process data sources of other modalities.
As shown in Figure \ref{join-example}, a multi-modal join operator, for example, allows for the joining of the patient table directly with the linked textual patient reports using SQL.
As such, the data analyst can correlate the relationship between the patient's age and their diagnoses in a simple and efficient manner without extracting a table from the documents in the first place.

The key idea behind MMOps such as the multi-modal join is that they can accept data sources of other modalities as input and produce tables as an output.
To that end, MMOps nicely integrate with the existing query processing capabilities of a traditional database system since MMOps can be composed in query plans along with other relational operators to enable complex analytical queries.
For example, after the multi-modal join operator shown in Figure \ref{join-example}, other relational operators such as a projection or a filter can be applied to provide rich query functionality to users.
Moreover, as we elaborate later in the paper, in addition to multi-modal joins, we also envision that MMDBs implement a wide spectrum of different MMOps including multi-modal scans, unions, and aggregations.

\noindent\textbf{Using Large Pre-trained Models.} 
Realizing such MMOps that can deal with modalities other than tables in a robust manner is far from trivial since other modalities are much harder to process.
For example, extracting structured (tabular) data from texts is a non-trivial task since the information can be found in various places in the source documents (e.g., information about the patient's diagnosis) or a different number of tuples need to be extracted per document (e.g., a report might contain two instead of only one diagnosis). 
To realize MMOps that can robustly deal with modalities such as texts, we propose to build on the advances of large pre-trained models such as GPT-3\cite{gpt3}.
While such models have been used for other complex data management tasks such as data deduplication or value imputation, they have not been used so far to implement query operators such as joins that can not only reason over tables but also over other modalities such as text or images.

\noindent\textbf{MMDBs for Tables and Text.} As a concrete contribution, in this paper, we present a first prototype of an MMDB.
In particular, with our MMDB prototype, we focus on text as an additional modality to tables by making use of pre-trained models for language.
The main idea is that we realize MMOPs such as multi-modal joins as downstream tasks on top of a pre-trained language model.
We believe that the ideas presented in this paper also transfer to other modalities such as images, audio, etc. where similar pre-trained models exist and thus MMOps for those modalities could be realized in a similar manner.
However, showing this is beyond the scope of this paper and presents an interesting avenue for future work.

For realizing MMOps, many different non-trivial challenges need to be tackled. 
First, while large pre-trained models for text such as GPT-3 have been used for  data processing tasks such as data cleaning and wrangling \cite{prefixtuning,plms4datawrangling}, it is not clear how multi-modal operations such as joins can be realized based on those models.
Hence, as a first direction in this paper, we propose a new pre-trained model that is based on a large language model which allows an MMDB to efficiently realize MMOps. 
Second, users of traditional database systems demand that query results be computed efficiently.
However, large pre-trained models for text such as GPT-3 are computationally intensive, making them costly to run in a database context.
Hence, as a direction we propose several optimization strategies to reduce the amount of processing necessary for executing queries over textual data.

\noindent\textbf{Contributions and Outline.}  To summarize, in this paper we present three major contributions to enable MMOps:
(1) As the main contribution, we present the MMDB-Model that is based on a pre-trained language model \cite{tabert}. For realizing the MMDB-Model, we provide  several important extensions to standard language models; i.e., a new pre-training procedure as well as a set of table-specific decoders to turn texts accurately into table data.
(2) As a second contribution, we show how MMOps can be realized on top of such an MMDB-Model. In particular, we first show a multi-modal scan as a core multi-modal operator of an MMDB, that can turn text collections into tables. Moreover, we discuss also more complex multi-modal operations like joins as shown before, as well as other operators including unions, and aggregations over text data. 
(3) We present several optimizations for query execution to make MMOps more efficient on large text collections. In particular, we present multi-modal materialized views and multi-modal (secondary) indexes as two simple yet effective techniques.

In the remainder, we first provide an overview of our MMDB system in Section \ref{sec:overview}.
Section \ref{sec:scan} then discusses how the multi-modal scan  is realized before Section \ref{sec:pre-training}  describes the details of the MMDB-Model.
Afterward, Section \ref{sec:other-ops} introduces the complex multi-modal operations and Section \ref{sec:performance-optimizations} presents our optimizations to tackle the challenges concerning efficiency.
Finally, we present an extensive experimental evaluation in Section \ref{sec:eval}, as well as related work in Section \ref{sec:related-work}, and conclude in Section \ref{sec:conclusion}. \begin{figure}
  \centering
  \includegraphics[width=1.0\linewidth]{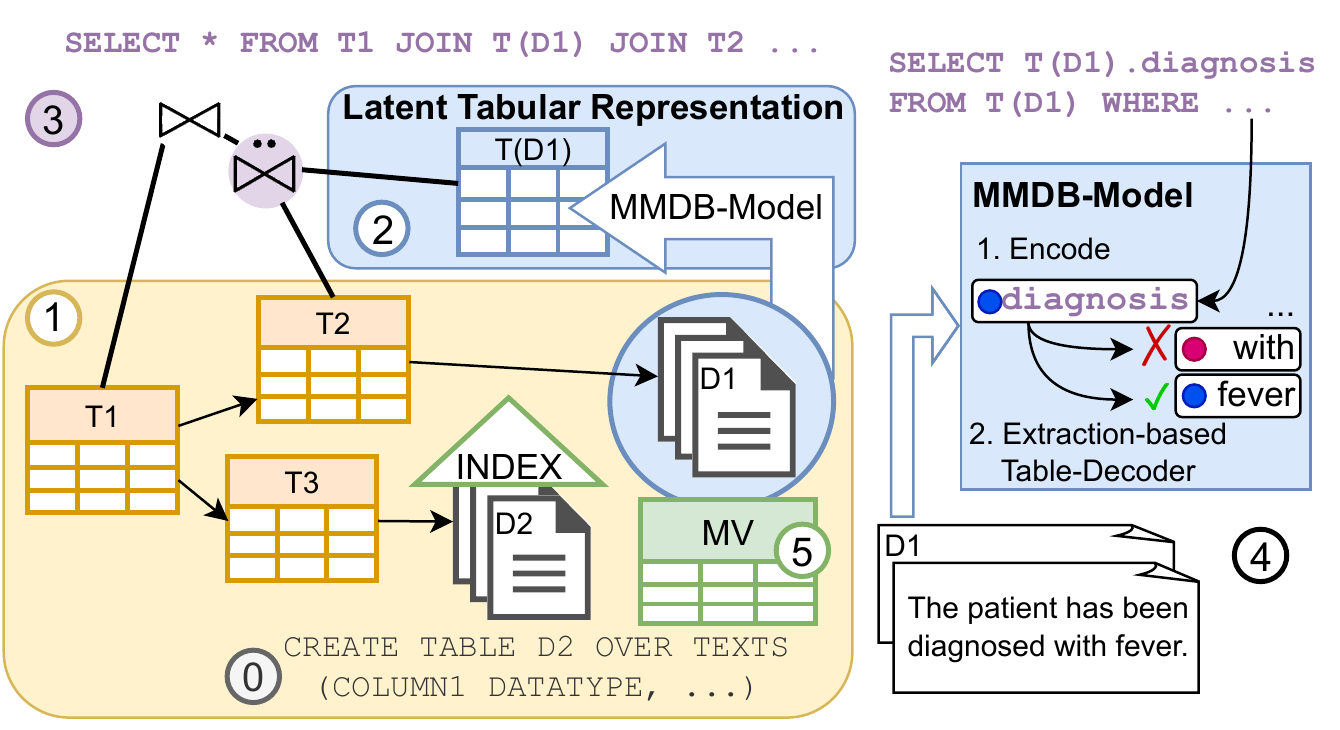}
  \vspace{-1.5ex}
  \caption{Overview of the MMDB Architecture: With our MMDB, \circled{0} Users can register text collections as tables using a \texttt{create table over text} statement by providing column names and datatypes.
  Such text collections are being made available for seamless querying as tables \circled{1}.
  This is enabled by the MMDB-Model which learns to create a tabular representation $T(D)$ of a text collection $D$ \circled{2}.
  \circled{3} For querying an MMDB, users can issue a multi-modal SQL query that is translated into a query plan that contains traditional and multi-modal database operators such as multi-modal joins or scans.
  \circled{4} To compute a multi-modal operation using the MMDB-Model, representations of query attributes and texts are computed first. Afterward, the MMDB-Model uses these representations to create the data for the output table by extracting values from the text.
  \circled{5} Multi-modal materialized views and multi-modal (secondary) indexes enable efficient query execution. }
  \vspace{-2.5ex}
  \label{fig:overview}
\end{figure}

\section{System Overview} \label{sec:overview}

In the following, we first discuss the key components of an MMDB before we explain the key contribution in contrast to existing approaches in more detail.

\subsection{The MMDB Architecture}

As depicted in Figure \ref{fig:overview}, MMDB comprises several key components to extend a classical database system and thus enable the seamless processing of text collections as tables.

\noindent\textbf{Multi-modal Database Storage.}
As a first key component, MMDB introduces an extension to the conventional database data storage that allows the integration of text collections as a first-class citizen (see \circled{1} in  Figure \ref{fig:overview}).
The main idea in an MMDB is that the text collections can be seamlessly treated as tables.
For adding a text collection as a table, only the schema of the queryable attributes needs to be specified (see \circled{0} in  Figure \ref{fig:overview}).
However, it is important that there is no need to tediously extract table data from text manually by using traditional techniques like RegEx or NER.

Instead, the transformation happens seamlessly using the MMDB-Model (see \circled{2} in  Figure \ref{fig:overview}).
The key novelty of realizing the MMDB-Model is that we teach the model the specific skills necessary to extract structured data from text collections given the schema of the table.
As we show in our evaluation, an MMDB can thus often be used out-of-the-box in a zero-shot mode, or in a few-shot mode by fine-tuning the MMDB-Model with only a few examples on a new unseen text collection.
We explain more details about the MMDB-Model later in Sections  \ref{sec:scan} to \ref{sec:pre-training}.

\noindent\textbf{Multi-modal Query Execution.}
For query execution over an MMDB, users can issue multi-modal SQL queries (see  \circled{3} in Figure \ref{fig:overview}).
Such multi-modal SQL queries are mapped to multi-modal query plans containing classical relational and so-called multi-modal operators (MMOps) such as multi-modal joins.
The most basic MMOp available in our MMDB is a multi-modal scan that produces a table from a text collection (see \circled{4} in Figure \ref{fig:overview}).
In its simplest form, the multi-modal scan extracts one row per text document.
In the following, we sketch how this operator can be realized by using our MMDB-Model.

For realizing a simple multi-modal scan that turns a text collection into a table, the operator feeds the attributes to be scanned (i.e., diagnosis) together with the input text of every document one-by-one into the MMDB-Model.
Afterward, the table encoder maps the text and query attributes into a joint latent space and the decoder of our MMDB-Model identifies spans of texts that qualify as values for the queried attributes (e.g., the text span in the input for filling the value of the query attribute diagnosis).
This extractive approach to derive table data from text has many advantages such as it avoids hallucination as we explain below.
Moreover, more complex multi-modal operators such as multi-modal scans that need to extract multiple rows per text document can be realized in a similar style as we discuss in Sections \ref{sec:scan} and \ref{sec:other-ops}.

To summarize, realizing the extraction of table data from text using MMOps has many advantages. 
First, the user can ad-hoc register and query new text collections without the need to first materialize a potentially large table.
This is in particular interesting if only a few documents of a large collection are queried by users.
However, materializing the data of an extraction is also possible in our MMDB using multi-modal materialized views (as we discuss below) which speed up queries that require the full table data. 
Second, in contrast to simple table extractions from text, we also provide other MMOps, such as multi-modal joins of a table and a text collection that can make use of the table data to provide better extractions as we discuss later.

\noindent\textbf{MMDB Optimizations.} Using language models at runtime to implement query operators incurs high computational overhead since these models typically have a high number of parameters. In order to reduce the computational overhead, in our MMDB, we employ two crucial optimizations to enable highly efficient multi-modal queries \circled{5}.
Multi-modal materialized views shift costly text processing to an offline extraction phase and multi-modal (secondary) indexes allow for efficient execution of online multi-modal queries that only need to read a fraction of the text collections.

\subsection{Discussion}\label{sec:discussion}

While there exist already approaches that transform textual data to tables like text-to-table \cite{text-to-table} or STable \cite{stable}, we think that these approaches are not suitable for implementing multi-modal database operations in an MMDB.

One key difference of our approach compared to text-to-table and STable is that the underlying models have to be trained from scratch for every new text collection.
While text-to-table, for example, also uses pre-trained language models as a basis, it is directly initialized with the pre-trained weights of a language model (i.e., BART \cite{bart}), which has been pre-trained on plain text only.
Different from that, we carefully design a new pre-training procedure that allows our MMDB-Model to provide more meaningful representations for table extraction and thus realize MMOps on unseen texts with just a few examples.

Moreover, another key difference is that text-to-table or STable produce the data for an output table using a transformer-based generative decoder \cite{transformers}.
The first disadvantage here is that the model can ``make up'' values, that are not actually present in the input text, but that the model picked up during (pre-)training.
This phenomenon is called hallucination \cite{hallucination}.
On top of that, transformer-based decoders output tables token-by-token in an autoregressive manner, which requires a pass through the decoder for every token in the output table.
This results in a computationally expensive decoding process as we show in our evaluation.
Finally, text-to-table or STable also do not directly support complex MMOps like joins, unions, and aggregations which allow an MMDB to produce higher-quality extractions as we explain later.

\section{Multi-modal Scan} \label{sec:scan}
\label{sec:scan}

In this section, we focus on the multi-modal scan which is a core multi-modal operator in an MMDB that takes a text collection and transforms it into a table using the MMDB-Model as shown in Figure \ref{fig:overview}  \circled{4}.
More complex operations such as joins and unions build on the ideas of the scan as we explain in Section \ref{sec:other-ops}.

\subsection{Basic and Complex Multi-modal Scan}
The task of a multi-modal scan is to turn a text collection $D$ into a table $T(D)$.
In a \emph{basic multi-modal scan},  each text $d \in D$ corresponds to a single tuple $t_d \in T(D)$.
For instance, a patient report might correspond to a single row in the examination table.
However, text collections are often more intricate, leading to the concept of \emph{complex multi-modal scans}.
We differentiate between complex multi-modal scans for multi-row texts and multi-table texts:

\noindent\textbf{Multi-row Texts.} In a multi-row text, a single text $d\in D$ may correspond to multiple tuples $t_d^1, t_d^2, \dots$ within the table $T(D)$.
As an example, a patient report documenting multiple examinations of the patient would result in multiple rows within an \texttt{examination} table.
To accommodate this complexity, a multi-modal scan must possess the capability to produce multiple tuples in response to a single input text.

\noindent\textbf{Multi-table Texts.}
In multi-table texts, the information within a text collection may be more complex and might be better represented by multiple extracted tables instead of one table.
For instance, the text collection of patient reports might consist of a table representing information of the examining physician as well as a additional tables capturing the information of the examination.

\noindent\textbf{Table Definition in an MMDB.} For querying text collections in our MMDB, a user needs to specify during table creation  (see \circled{0} in  Figure \ref{fig:overview}) of  which table type a text collection is; i.e., if the texts are single-row texts, multi-row texts, or multi-table texts. 
Moreover, for multi-row texts, a so-called set of \emph{identifying attributes} (i.e., an attribute where values are unique within a document) needs to be defined in the table schema when creating a table for a text collection, which is used for the multi-row scans as explained later in Section \ref{sec:scan:complex}.
For instance, for examination reports, the column \texttt{patient name} would be an identifying attribute, because it identifies different patients mentioned in the same text.

\subsection{The Basic Multi-Modal Scan}
\label{sec:scan:basic}

In the following, we first focus on how to use our MMDB-Model for the basic multi-modal scan which transforms one text into one row.
Afterward, we explain how we can use the MMDB-Model for the more complex scenarios (multi-row and multi-table texts).
Similar to other approaches for table extraction from text \cite{text-to-table, stable}, we employ an encoder-decoder architecture for our model.
In particular, for realizing our MMDB-Model we built on TaBERT \cite{tabert} as the encoder, which is a pre-trained language model that is already aware of tabular data.

However, in order to support MMOps as downstream tasks several significant extensions are needed as we discuss below.
For example, different from other existing approaches, our MMDB-Model comes with a set of new decoder heads that are not transformer-based and thus eliminate hallucination and are more efficient as discussed in Section \ref{sec:discussion}.
Moreover, the encoder and decoder are pre-trained by tasks that learn in general to extract table data from text as presented in Section \ref{sec:pre-training}. This allows the usage of the model in zero and few-shot settings where only a few example extractions per table are needed to fine-tune the model on new unseen text collections.

\begin{figure}[t]
  \centering
  \includegraphics[width=1.0\linewidth]{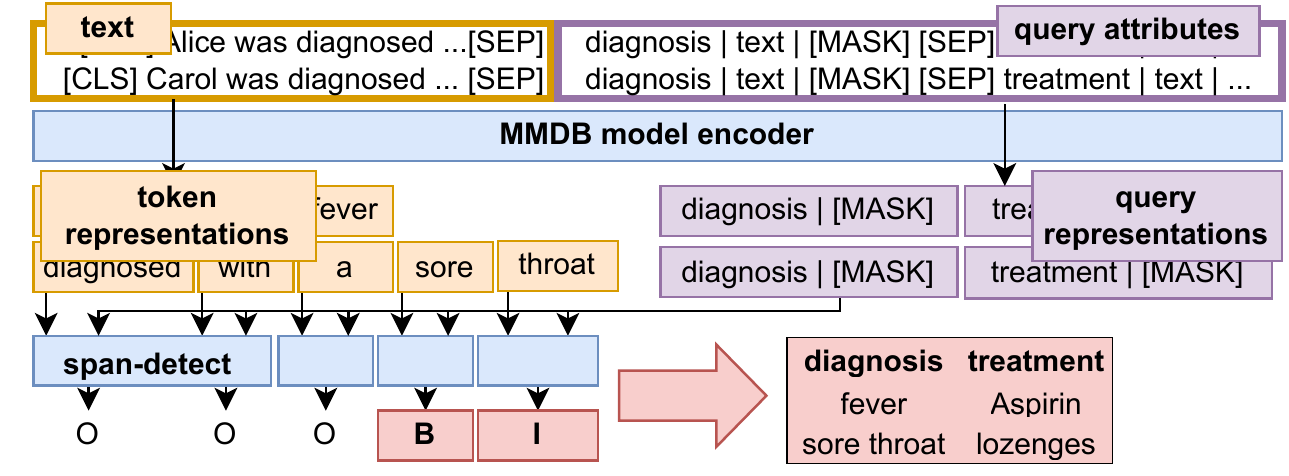}
  \vspace{-4.5ex}
  \caption{Overview of a basic scan. The scan transforms each input text into a single row. As input (at the top), each text is paired with the query attributes using a MASK token for the values to be extracted. The MMDB-Model computes representations of all text tokens and the attribute information which include the MASK tokens. Based on these representations, the span-detect head assigns I-O-B tags for the text tokens to be extracted for each queried attribute. In this case \emph{sore throat} is extracted for the diagnosis. Finally, the result table is constructed from the extracted text passages.}
  \vspace{-2.5ex}
  \label{fig:basic}
\end{figure}

\noindent\textbf{Encoder.} The process of computing a multi-modal scan operation using the MMDB-Model is sketched in Figure \ref{fig:basic}.
In a nutshell, the encoder first maps the input text into a latent space by computing vector representations $\hat w$ for all input tokens $w$.
The decoder then utilizes these representations to identify all relevant tokens for the input query, similar to how extractive question answering is solved using language models \cite{bert}.
However, transforming texts to tables is much more complex than extractive question answering since it involves the extraction of potentially many different values of one row or the removal of duplicates as we discuss below.
To support the extraction of values for multiple columns using our table decoder, we encode all the columns jointly with the input texts as shown in Figure \ref{fig:basic}.
The columns are encoded using column names and types. The MASK token for a query attribute indicates that the value of the attribute needs to be extracted from the text. For other operations such as joins some attribute values will be additionally provided as input as we discuss later. 

\noindent\textbf{Decoder.} Our extraction-based table decoder consists of several classification heads that take on different tasks for table decoding.
The most important component is the span-detect (SD) head that extracts values, potentially consisting of multiple tokens, for all queried columns.
In essence, the layers of the SD head pair the representation of each token of the text input $\hat w$ with the representation of each masked cell $\hat c$ for the column to be extracted and classify whether the token is part of a value for the column.
We employ so-called \emph{I-O-B (Inside-Outside-Beginning) tags} to extract potentially multiple tokens for each column (i.e., complete text spans). 
With I-O-B tags, the first text token for a value that should be extracted is marked with a B-tag to indicate its beginning, while subsequent tokens belonging to the same value receive an I-tag to indicate they are inside the value.
Tokens that do not belong to a value are marked with an O-tag.
This clearly allows us to extract a single table row for each input text, as shown in Figure \ref{fig:basic}.

The I-O-B tags are predicted by the SD head.
To realize the SD head, we use two learned matrices $W^{SD}_B$ and $W^{SD}_I$ --- each representing a single additional layer after the encoder --- and a learned threshold $thresh^{SD}_O$.
The tag, for each pair of token- and cell-representation $(\hat w, \hat c)$, is calculated  based on these learned matrices as follows:

\vspace{-2.5ex}
\begin{equation}
\label{eqn:iob-tagging}
tag \in \argmax_{x\in\{I, O, B\}}
\begin{cases}
\hat w^T \cdot W_x^{SD} \cdot \hat c & if x \in \{B, I\}\\
thresh_O^{SD} & else
\end{cases}
\end{equation}

\subsection{Multi-row and Multi-table Texts} 
\label{sec:scan:complex}

Multi-row texts and multi-table texts present a more complex scenario as discussed before.
In the following, we discuss both cases in detail.

\begin{figure}[t]
  \centering
  \includegraphics[width=1.0\linewidth]{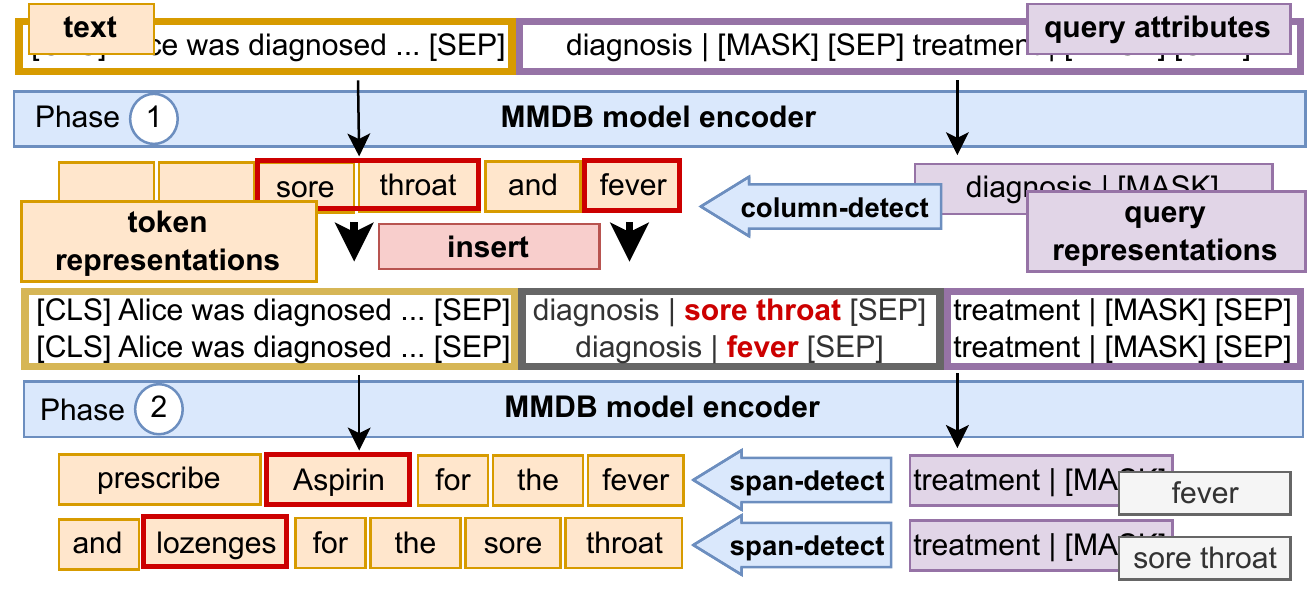}
  \vspace{-4.5ex}
  \caption{Overview of a complex multi-modal scan. To extract an arbitrary number of tuples, the input sequence is passed multiple times through the encoder. The input for extracting one tuple (as shown at the top) consists of the input text paired with the column names to be extracted. In phase \circled{1}, values of the so-called \emph{identifying attribute} (e.g., \texttt{diagnosis}) are extracted, using the column-detect head of our decoder. In phase \circled{2}, the model computes the values for all other columns. For example, depending on the \texttt{diagnosis}, the \texttt{treatments} (e.g., Aspirin for fever) are extracted.}
  \vspace{-2.5ex}
  \label{fig:complex}
\end{figure}

\noindent\textbf{Multi-row Texts.} In order to generate an arbitrary number of tuples per document, we utilize an iterative approach as depicted in Figure \ref{fig:complex} and formalized in Algorithm \ref{alg:complex} for a single text $d \in D$.
For the purpose of simplification but without loss of generality, we make the assumption that there exists a single so-called identifying attribute $c_{key}$ in the table definition of $D$ that is defined in the table schema of the text collection as discussed in Section \ref{sec:overview}. 
In our example scenario, we assume the attribute $c_{key}=$\texttt{diagnosis} serves as the identifying attribute.

In the first phase of the procedure, the model extracts all values for the identifying attribute $c_{key}$, using a so-called column-detect head in the decoder.
The number of values extracted in this step determines the number of tuples that will be generated for a text collection.
As demonstrated in Figure \ref{fig:complex}, the names of the diagnoses, such as \emph{sore throat} and \emph{fever}, are extracted as values for the identifying attribute, and thus two output rows will be generated.
Different from before, we do not use the span-detect head to extract values for $c_{key}$, but a column-detect (CD) head that is conceptually identical to the span-detect head but consists of a different set of weights $W_B^{CD}$, $W_I^{CD}$ and threshold $thresh_O^{CD}$.
The reason for this is that the CD-head is pre-trained on different tasks to extract values for multiple entities and not only one, as we will discuss in Section \ref{sec:pre-training}.
In our ablation study, we show that this results in more accurate extractions for some datasets.

In the second phase, the extraction process is conducted on the remaining  columns that depend on the identifying attribute, denoted as $c_1, c_2, \dots$.
To accomplish this, the MASK token of the identifying attribute $c_{key}$ is now replaced with the values extracted in the first phase, as depicted in Figure \ref{fig:complex}.
These sequences are then processed by the model to extract the values for the dependent columns, using the span-detect head to only extract the values corresponding to the value extracted in the first phase.
This iterative approach enables the model to effectively process complex scans using only a few passes through the transformer per input text.
As we will show in our evaluation, this procedure is computationally much less expensive than using a transformer-based decoder that is used by existing approaches such as text-to-table \cite{text-to-table}.

Finally, we want to mention that our approach for multi-row scans can easily be extended for the case where the identifying attributes is composed of multiple columns.
For instance, in a text collection of medical records, the uniqueness of a \texttt{diagnosis} may be determined by both its name and the \texttt{date} of \texttt{diagnosis} (i.e., the generated table should contain multiple tuples for the diagnosis \emph{fever}, if it has been diagnosed on multiple dates).
In this case, we first extract all \texttt{diagnoses} using the column-detect head, then we collect all corresponding \texttt{dates} using the span-detect head.
Afterward, the values for the remaining (dependent) query attributes are extracted for each pair of \texttt{diagnosis} and \texttt{date} like before.

\noindent\textbf{Multi-table Texts.}  Multi-table texts do not require additional algorithmic considerations.
In this case, we simply prepend the name of the table we want to extract from the text to the name of the identifying attribute $c_{key}$.
The model is trained to infer from the column names which table to extract.

\begin{algorithm}
\caption{Complex multi-modal scan}\label{alg:complex}
\begin{algorithmic}
\Require $d=(w_1, ..., w_n)$, $C=(c_{key}, c_1, ..., c_m)$ \\
\textbf{1. Compute representations}  \Comment{1st phase}
\State $(\hat{w_1}, ..., \hat{w_n}), (\hat{c_{key}}, \hat{c_1}, ..., \hat{c_m}) \gets encoder(d, C)$ \\
\textbf{2. Find values} for $c_{key}$
\State $V_{key} \gets column-detect((\hat{w_1}, ..., \hat{w_n}), \hat{c_{key}})$
\State $\{v_1^{key}, ..., v_o^{key}\} \gets duplicate-detect(V_{key})$ \\
\textbf{3. Get values for $c_1, ..., c_m$}
\For{$v_i^{key} \in \{v_1^{key}, ..., v_o^{key}\}$} \Comment{2nd phase}
\State $(\hat{w_1^i}, ..., \hat{w_n^i}), (\hat{c_1^i}, ..., \hat{c_m^i}) \gets encoder(d, v_i^{key}, c_1, ..., c_m)$
\State $(V_1^{i}, ..., V_m^{i}) \gets span-detect((\hat{w_1^i}, ..., \hat{w_n^i}), (\hat{c_1^i}, ..., \hat{c_m^i}))$
\State $(\{v_1^{i}\}, ..., \{v_m^{i}\}) \gets duplicate-detect(V_1^{i}, ..., V_m^{i})$
\EndFor \\
\Return $\begin{bmatrix}
v_1^{key} & v_1^1 & \dots & v_m^1\\
\vdots & \vdots & \ddots & \vdots \\
v_o^{key} & v_1^o & \dots & v_m^o
\end{bmatrix}$

\end{algorithmic}
\end{algorithm}

\subsection{Removing Duplicates}
In the case of multi-row texts, even when using the concept of identifying attributes, duplicate rows can occur due to different spellings or synonyms.
For example, the same text might contain the value \emph{fever} and \emph{high body temperature} which refers to the same diagnosis.
In such cases, a multi-modal scan might extract two rows for the same diagnosis. Hence, we need to remove such duplicates in a multi-modal scan after the extraction.
However, separating cases of duplicates from non-duplicates is non-trivial.

To address this issue, we thus use an additional duplicate-detect head as part of our decoder which allows us to filter out duplicates.
The duplicate-detect head detects duplicate values based on their representation in the latent space.
To compute a representation of a value that potentially consists of multiple tokens, we employ span-representations as introduced by \citet{span-representation}.
The duplicate-detect head consists, similar to the span-detect head, of a learned matrix $W^{DD}$ and a learned threshold $thresh^{DD}$.
Two values for a column are considered duplicates if the learned similarity of the span representations $s_A$ and $s_B$ is larger than the threshold $s_A^T \cdot W^{DD} \cdot s_B > thresh{DD}$.
 
\begin{figure*}[t]
  \centering
  \includegraphics[width=\linewidth]{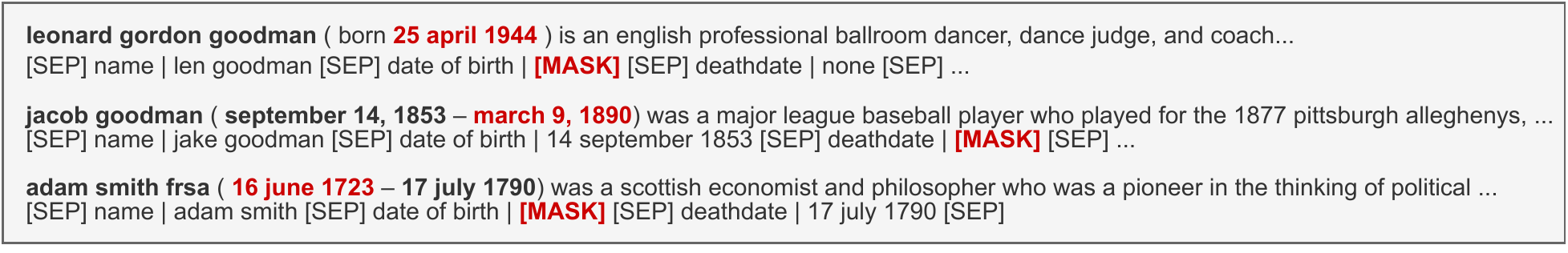}
  \vspace{-6.5ex}
  \caption{Example pre-training sample consisting of three rows and texts from our pre-training dataset. Wikipedia abstracts are paired with rows constructed from Wikdata containing information about the same entities. Labels for the MCR objective are marked in red and labels for the CTA objective are in bold. }
  \label{fig:corpus-example}
  \vspace{-2.5ex}
\end{figure*}

\section{The MMDB-Model}
\label{sec:pre-training}

At the core of the multi-modal operators such as the multi-modal scan introduced before is the MMDB-Model.
An important aspect is that the MMDB-Model does not require extensive training data for every new set of documents that should be added to the MMDB.
To enable this, we pre-train the MMDB-Model with new objectives such that the model learns the general skills to extract tables from texts for unseen documents.
As a result, as we show in our evaluation, the MMDB-Model can be used in zero- or few-shot mode with high accuracy, which is in stark contrast to existing approaches such as text-to-table \cite{text-to-table}.

\subsection{Pre-training Objectives}
The pre-training procedure is comprised of three new pre-training objectives, each targeting different skills necessary to perform table extraction for multi-modal database operators.

\vspace{0.5ex}\noindent\textbf{Skill 1 - Extract values for specified columns.}
Multi-modal database operators extract values for specified columns from text.
For example, in a scan for multi-row texts, different values for the same column (e.g., diagnosis) need to be extracted. 
To support that our model learns to detect values for certain columns, we introduce the \emph{Column-Text-Alignment (CTA)} task, which aligns table columns to text.
In this task, we pair texts with tables and compute a representation for all columns using the schema information (attribute names and types) and the actual values of the column.
Afterward, we let the model detect segments of text that are a potential value for a column.
This task not only enables efficient multi-modal scans that we introduced before, but also more complex operations such as a multi-modal union or join as we explain later in Section \ref{sec:other-ops}.

\vspace{0.5ex}\noindent\textbf{Skill 2 - Extract values for specified rows.}
Second, it is often important for the model to align texts with table rows and not only columns. For example, in a multi-modal join only the values (e.g., diagnosis) corresponding to a linked table row (e.g., a patient) should be extracted.
To let our model learn connections between text passages and table rows, we introduce an additional \emph{Masked Cell Reconstruction (MCR)} pre-training task.
In this task, we pair table rows and texts and mask out random cell values of the table row (that occur in both text and table row), and ask the model to reconstruct the masked values from text using the signals from neighboring cells, including the cells from the same row.
Different to CTA, MCR thus teaches the model to extract values for a certain row only (e.g. only the diagnoses of Bob), while CTA extracts all values corresponding to a column (e.g. all diagnoses mentioned in a text). 

\vspace{0.5ex}\noindent\textbf{Skill 3 - Detect duplicates.} As discussed before, texts often contain the same information multiple times. 
For example, a patient report might mention twice that a patient was diagnosed with fever in different places in the document, potentially using different spellings or synonyms.
As such, it is important to detect when two extracted spans refer to the same or to different entities to avoid unwanted duplicates in the output of multi-modal operators.
Therefore, we introduce the \emph{Duplicate-Detection (DD)} pre-training task as our final pre-training objective.
For each column, \emph{DD} takes pairs of spans of values of the text as input and is trained to predict whether they are the same or not.
For this objective, our training corpora that we explain below leverages T-REx \cite{trex} which contains  information about which spans in a text relate to the same actual entity.

\subsection{Pre-training Details}
\label{sec:pretraining:details}

In the following, we discuss the details of our pre-training objectives discussed before.
In a nutshell, for pre-training we pair the transformer-based encoder with our table-decoder, and train both end-to-end.
For pre-training the MMDB-Model, we feed pairs of texts and a sample of table rows into the encoder of our model and apply the following pre-training objectives. 
Our table encoder makes use of vertical self attention as introduced in \cite{tabert} to let signal flow between the individual linearized rows.
Moreover, remember that we use a multi-head table decoder with the the span-detect, the column-detect and the duplicate-detect head as already introduced in Section \ref{sec:scan}.

\vspace{0.5ex}\noindent\textbf{MCR and CTA Objectives.} Let $\hat w_{i, 1}, \hat w_{i, 2}, ..., \hat w_{i, n}$ be the token representations of the $i$-th text and let $\hat c_{i, 1}, \hat c_{i, 2}, ..., \hat c_{i, m}$ be the cell representations of the $i$-th tuple, after encoding the table row with our encoder.
For pre-training the MCR and CTA objectives as introduced before, we use the matrices $W_B$, $W_I$ and the threshold $thresh_O$ from the span-detect and column-detect heads.

For MCR, we use the cell representations of masked cells $\hat c_{i, j}$ and the token representations $\hat w_{i, 1}, \hat w_{i, 2}, ..., \hat w_{i, n}$ for the extractions.
For CTA, we use the column representation $\hat h_{j} = \frac{1}{k}\sum_{l = 1}^{l \le k}\hat c_{l, j}$ as input instead of the cell representation of the masked cell.
We train the matrices and the threshold using cross-entropy to obtain loss values $\mathcal{L}_{MCR}$ and $\mathcal{L}_{CTA}$.

To ensure that the model utilizes the actual text values during pre-training and not only the schema information (i.e., column names) of the table we aim to extract, we randomly mask out also column names from the linearized input to our model in 15\% of the cases.

\vspace{0.5ex}\noindent\textbf{DD Objective.} For the DD pre-training objective, we use the duplicate-detect head consisting of the learned matrix $W_{DD}$ and threshold $thresh_{DD}$, as introduced in Section \ref{sec:scan}.
We train the matrix and the threshold by minimizing a cross-entropy loss $\mathcal{L}_{DD}$.

\vspace{0.5ex}\noindent\textbf{Combined Pre-Training.} For the pre-training, we use a combined pre-training as suggested in \cite{bert, strug} where we add up all the losses of all objectives and train the entire model (including the encoder) using this multi-task loss.
To balance the losses we found it beneficial to use a weighted sum. We choose $\alpha=300$, $\beta=80$, $\gamma=\delta=1$ by examining the performance on a development set.
\begin{equation}
    \mathcal{L} = \alpha\cdot \mathcal{L}_{MCR} + \beta\cdot\mathcal{L}_{CTA} + \gamma\cdot\mathcal{L}_{DD} + \delta\cdot\mathcal{L}_{MLM} 
\end{equation}

Moreover, for the pre-training, we realized that our model benefits from adding the original Masked Language Model loss $\mathcal{L}_{MLM}$ of BERT.
We thus added $\mathcal{L}_{MLM}$ to the other losses as well.
We compare the effect of the different pre-training objectives in our ablation study in Section \ref{sec:ablation}.

\subsection{A New Pre-Training Corpus}\label{sec:corpus}

Unfortunately, currently no pre-training corpus exists that allows us to pre-train our MMDB-model as outlined in Section \ref{sec:pre-training}.
In contrast to existing corpora such as \cite{book-corpus,webtables,gittables}, we need a different parallel corpus where the texts contain the same information (e.g., same entities) as the tables, and where we know which cell values also occur in a text, allowing us to mask them for pre-training.
Hence, we constructed a new parallel open-domain pre-training corpus from Wikidata and Wikipedia for pre-training  multi-modal database models.
We will open-source the corpus together with our code once our paper is published.

The main idea of our dataset is that we make use of T-REx \cite{trex}, a large-scale alignment of Wikidata triples and Wikipedia abstracts.
The T-REx dataset contains 11 million alignments of Wikidata triples to Wikipedia abstracts.
All the 3.09 million abstracts occurring in T-REx are also part of our dataset.
T-REx itself is created automatically using the distant supervision assumption for computing the alignment and can thus be noisy sometimes, but it allows us to construct a large pre-training corpus with objectives aligned to the downstream task of multi-modal database operations.
T-REx has been used by other researchers for training and evaluating their machine learning models \cite{kilt}.

Hence, we use the alignment of T-REx as a starting point to construct our parallel corpus of texts and relational tables.
T-REx contains information about the location of mentioned entities in the texts, which we can use as labels for our pre-training objectives.
As texts, we simply use the aligned Wikipedia abstracts and construct additional tables using Wikidata, by grouping similar entities together in a table and using Wikidata properties as columns.
We use several techniques to obtain a rich and diverse dataset, e.g. we randomize column names using Wikidata's aliases or concatenate multiple abstracts to simulate texts about multiple entities.

 \section{Other Multi-modal Operators} \label{sec:other-ops}

The multi-modal scan operator ignores valuable information stored in the other tables of the database.
In the following, we discuss additional multi-modal operators that can incorporate information from other tables in an MMDB to improve the quality of extractions.

\subsection{Multi-modal Joins}
A multi-modal join computes a join over a table $T$ and a text collection $D$ . 
To be more precise, in an MMDB a document $d \in D$ is linked to (at least) one tuple $t \in T$ as a join partner.
During execution, the multi-modal join extracts query attributes from $d$ and appends them to $t$.
A naive version of a multi-modal join would be to extract these attributes similar to a multi-modal scan.

However, different from a multi-modal scan, with a multi-modal join, we can make use of the data from the tuple $t \in T$ as evidence for extracting attributes from texts.
For example, assume the user joins a patients table with a document collection of patient reports where one report might contain texts for multiple patients.
Hence, if the name of the patient in $t$ and the name in the document $d$ match, this is a strong signal that the surrounding text in the document is about the requested patient and therefore relevant for extraction.

\begin{figure}[t]
  \centering
  \includegraphics[width=1.0\linewidth]{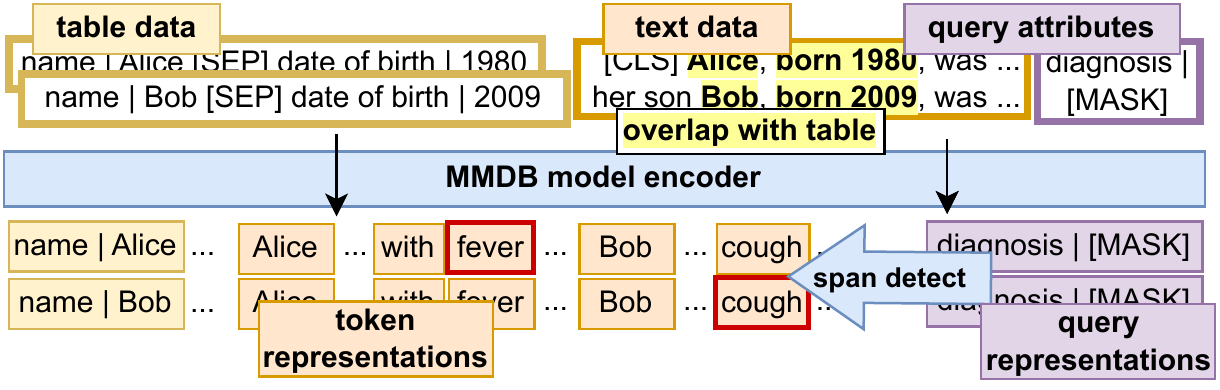}
  \vspace{-4.5ex}
  \caption{A multi-modal join uses the tuples of the join partner (top left, in yellow) as additional evidence to improve extractions. Some values (name, date of birth) are present in both the text and the join partner and help the model to locate relevant passages of text. We pair each text with their linked table rows and extract a tuple for each linked row.}
    \vspace{-4.5ex}
  \label{fig:join}
\end{figure}

To incorporate the data from the join partner $t$ as additional evidence to extract values from $d$, we feed attributes of $t$ along with the text into the MMDB-Model.
To be more precise, an input sequence to the MMDB-Model consists of three parts: the text document $d$, the query attributes for extractions, and column-value pairs from the join partner $t$, as depicted in Figure \ref{fig:join}.
For the extraction, we use the span-detect head that we also use for extracting values for the scan.

\subsection{Multi-modal Unions}
Different from a multi-modal join, a multi-modal union extracts rows of a text collection to append them to rows of a table.
For a multi-modal union, evidence from the table can thus provide example rows that help with the the extraction.
For instance, if a patient stored in the table has \emph{fever} as a diagnosis, this can enrich the model's understanding in regard that it needs to extract diagnoses for illnesses and not for computer problems.

\begin{figure}[t]
  \centering
  \includegraphics[width=1.0\linewidth]{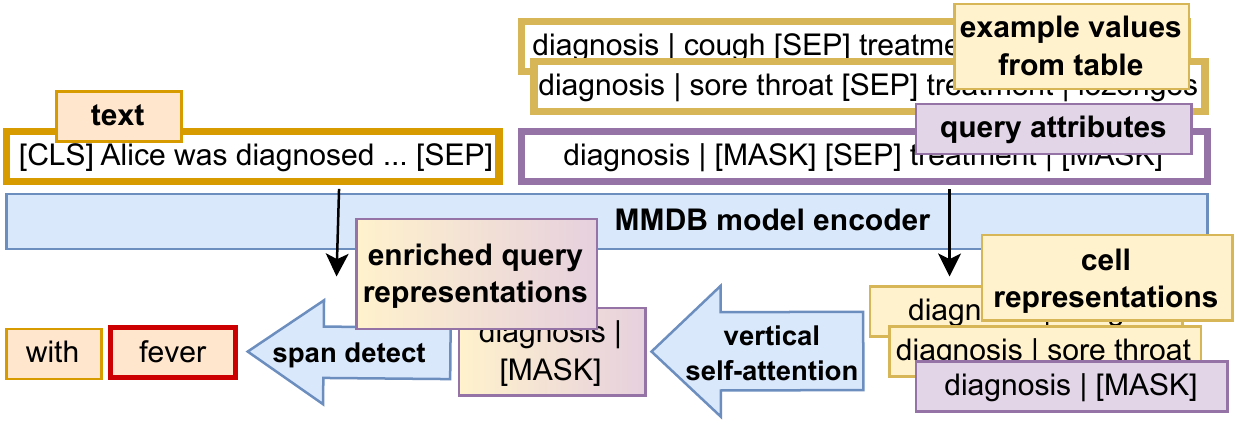}
  \vspace{-4.5ex}
  \caption{A multi-modal union uses the values from the table (top right, in yellow) as additional evidence to enrich the representations of the masked columns using vertical self-attention. These values act as example values for the columns and ultimately result in more accurate extractions.}
    \vspace{-4.5ex}
  \label{fig:union}
\end{figure}

To provide existing rows as example, we randomly sample a few rows from the table and encode them along with the attributes to be extracted from the texts using our encoder as shown in Figure \ref{fig:union}.
In order to let signal from example rows enrich our column representations, in our MMDB-Model we use vertical self-attention layers as mentioned before to teach the MMDB-Model to make use of example values for extracting rows from documents.
The vertical self-attention layer applies self-attention along columns of different input rows.

The input to the model consists of input sequences of multiple sample rows from the table and the input sequence containing text for which we aim to extract additional rows.
Based on this input, we compute the the result of the union similar to the multi-modal scan.
The only difference is that in the first phase of the multi-phase algorithm, we compute a column representation by taking the mean of the cell representations of the column values as well as the masked query attribute.
We then extract spans from texts using this column-representation.
For span-detect, we use the enriched representation of the masked column.

Finally, an important point is that we provide the same example rows from the table for all texts to be transformed since encoding additional example rows means a significant runtime overhead.
This way, the example rows need to be encoded only once for many texts which improves the efficiency of the union significantly.

\subsection{Multi-modal Aggregation}
Multi-modal aggregation is different from joins and unions since it does not make use of data in an existing table to enhance the extraction.
Instead, a multi-modal aggregation directly works on a collection of texts and allows the grouping/aggregation on columns extracted from the text (e.g., to count how often a certain diagnoses was found across all patients).
A naive solution to multi-modal aggregation would be to use a multi-modal scan and then group directly based on the extracted values.
However, again the variability of language can result in the same values being expressed in different ways in text.
To address this issue and to ensure that different surface forms (e.g., fever vs. high temperature) are placed in the same group, we aggregate values based on their representation, similar to how we detect duplicates.
As such, we re-use the duplicate-detect head introduced for de-duplication in scans to predict whether two values should belong to the same group.
 \section{Performance Optimizations} \label{sec:performance-optimizations}

Query latency is a major concern of database research for decades and is thus also an important aspect of MMDB.
At the same time, incorporating a language model in an MMDB, as proposed in this paper as well as other works \cite{neuraldb, symphony}, comes with significant computational overhead.
Since we do not want to sacrifice extraction quality by using subpar models, we thus present several optimizations to our system that allow us to use a language model efficiently.

\subsection{Multi-modal Materialized Views}
Materialized views are a well-established technique in databases for optimizing complex queries by performing computation during an offline phase.
In the context of MMDB, multi-modal materialized views aim to remove the computational burden of transforming documents into tables during query time.
By storing the result of a multi-modal operator, such as scan, join, or union, in a materialized result table, multi-modal materialized views function as a cache that avoids the need to use the language model at query execution.

One major difference between a multi-modal view and a classical materialized view is that we normalize the extracted result using our MMDB-Model.
In particular, we use the duplicate-detect head of the model to replace similar values that occur in different rows with a unified value if they have the same meaning.
For instance, values such as \emph{high body temperature} and \emph{fever} that are used for the same \emph{diagnosis} would be replaced by one normalized value (e.g., \emph{fever}), to be able to effectively aggregate on the \emph{diagnosis} column of the materialized view.

\subsection{Multi-modal Secondary Indexes}
While materialized views clearly speed-up queries, they have several (known) drawbacks.
These include the significant upfront cost of materialization and reduced flexibility, as they assume certain query patterns.
As an alternative to avoid materialization and still answer queries efficiently, we introduce secondary multi-modal indexes.
Multi-modal indexes are used in our MMDB when texts should be filtered by an attribute extracted from the text.
For example, users might be interested only in \texttt{treatments} for patients diagnosed with \emph{fever}.
In the current version of MMDB, in fact, we require the existence of such a secondary multi-modal index on an attribute if it should be used for filtering.

The core idea of a multi-modal secondary index is similar to a traditional index; i.e., the index maps a search key for the query attribute (e.g., diagnosis) to text documents that contain the search key.
For constructing a multi-modal secondary index for a query attribute, our approach leverages our MMDB-Model and the column-detect head to extract the values for the search key from all texts.
For building the index, we put the extracted values and the paths to the text documents into a traditional index such as a B-tree or a hash-index to be able to quickly retrieve texts at query time.
However, unlike traditional indexes, we group similar extracted values into one index entry using the duplicate-detect head of our MMDB-Model.
This allows the index to return text document containing the diagnosis \emph{high temperature}, even if the user queries for \emph{fever}.
Note, after the resulting texts are returned by an index lookup, the texts still need to be processed by a multi-modal operator that transforms the text in a row  and normalizes the output (e.g., for grouping and aggregation) as described before in Sections \ref{sec:scan} and \ref{sec:other-ops}.

 \section{Experimental Evaluation}
\label{sec:eval}

In this section, we present the results of our experimental evaluation which justifies the design of the multi-modal database (MMDB) and the MMDB-Model in particular.
For this, we compare our implementation of multi-modal database operations (MMOps) with an implementation using text-to-table\cite{text-to-table} and show that our model is both computationally more efficient and achieves better extractions with limited training data.

\noindent\textbf{Evaluation Datasets.}
For our evaluation, we use the \emph{rotowire} dataset \cite{rotowire}, which is a dataset that has been used in other papers.
The dataset consists of basketball game reports and tables summarizing the most important statistics of each game.
In the \emph{rotowire} dataset, each game report is paired with two tables: one containing statistics concerning the two opposing \texttt{teams} and one containing statistics about the participating \texttt{players}.
We use the post-processed dataset version provided by \citet{text-to-table} that filtered out table values not present in the text.
On this dataset, the goal of the multi-modal scan is to transform the game reports into tables containing the statistics.
Here, we are able to perform complex scan operations since multiple tuples and tables need to be extracted per input text.
To be able to simulate join operations, we combine the dataset with two tables containing general information about basketball \texttt{teams} and \texttt{players}, which we constructed from Wikipedia infoboxes.
We linked each game report to the tuples of \texttt{teams} and \texttt{players} of the general information table, that participated in the game.
The multi-modal join will extract statistics from the game reports and add these as columns to the general information table.
For the union operation, we union the game reports with the statistics tables of the training set.

Unfortunately, the other datasets used in \cite{text-to-table} are not suitable for our evaluation, because they are either a subset of our pre-training dataset \cite{wikibio}, or the contained texts are too simplistic and consist of only one or two sentences \cite{e2e}.
Hence, we construct an additional dataset using T-REx \cite{trex}, similar to how we constructed our pre-training dataset (see Section \ref{sec:corpus}).
Importantly, this dataset is comprised of three unseen domains from Wikipedia that we left out in our pre-training dataset: nobel prize winners, skyscrapers, and countries.
The dataset consists of the Wikipedia abstracts and corresponding table rows, which we split into train, test, and validation sets.
For this data, in contrast to the \emph{rotowire} dataset, only a single row needs to be extracted per input text.

\noindent\textbf{Baselines.}
We compare our MMDB which is based on the MMDB-Model to text-to-table \cite{text-to-table}, the state-of-the-art approach for translating texts to tables.
The more recent STable approach \cite{stable} was not able to match its result on text-to-table tasks, despite increased computational overhead, hence we omit a comparison.

\noindent\textbf{Experimental Setup.}
All experiments were executed on a DGX A100 server\footnote{https://images.nvidia.com/aem-dam/en-zz/Solutions/data-center/dgx-a100/dgxa100-system-architecture-white-paper.pdf}.
For pre-training, we used 4 GPUs which took approximately 8 days for training our model for 6 epochs on our pre-training dataset.
For fine-tuning and inference --- in particular, for the performance measurements in experiment 2 --- we used 2 GPUs only.

\subsection{Exp. 1: Accuracy of Multi-Modal Queries}

In this first experiment, we measure how accurately our model is able to answer queries containing MMOps in contrast to using text-to-table\cite{text-to-table}.
Since modern DBMS are used in various domains and collection of training data for these domains is often expensive, we focus on scenarios where only limited data is available for fine-tuning.
Hence, we train our model on small subsets of the training set, and do not use the validation set for hyperparameter tuning.
Instead, we use the hyperparameters obtained from optimization on a separate subset of T-REx.
For all operations, we compute an F1 score for the extracted table values for each text in the test set and report the mean of these F1 scores.

\begin{figure}
    \centering 
    \includegraphics[width=0.85\columnwidth]{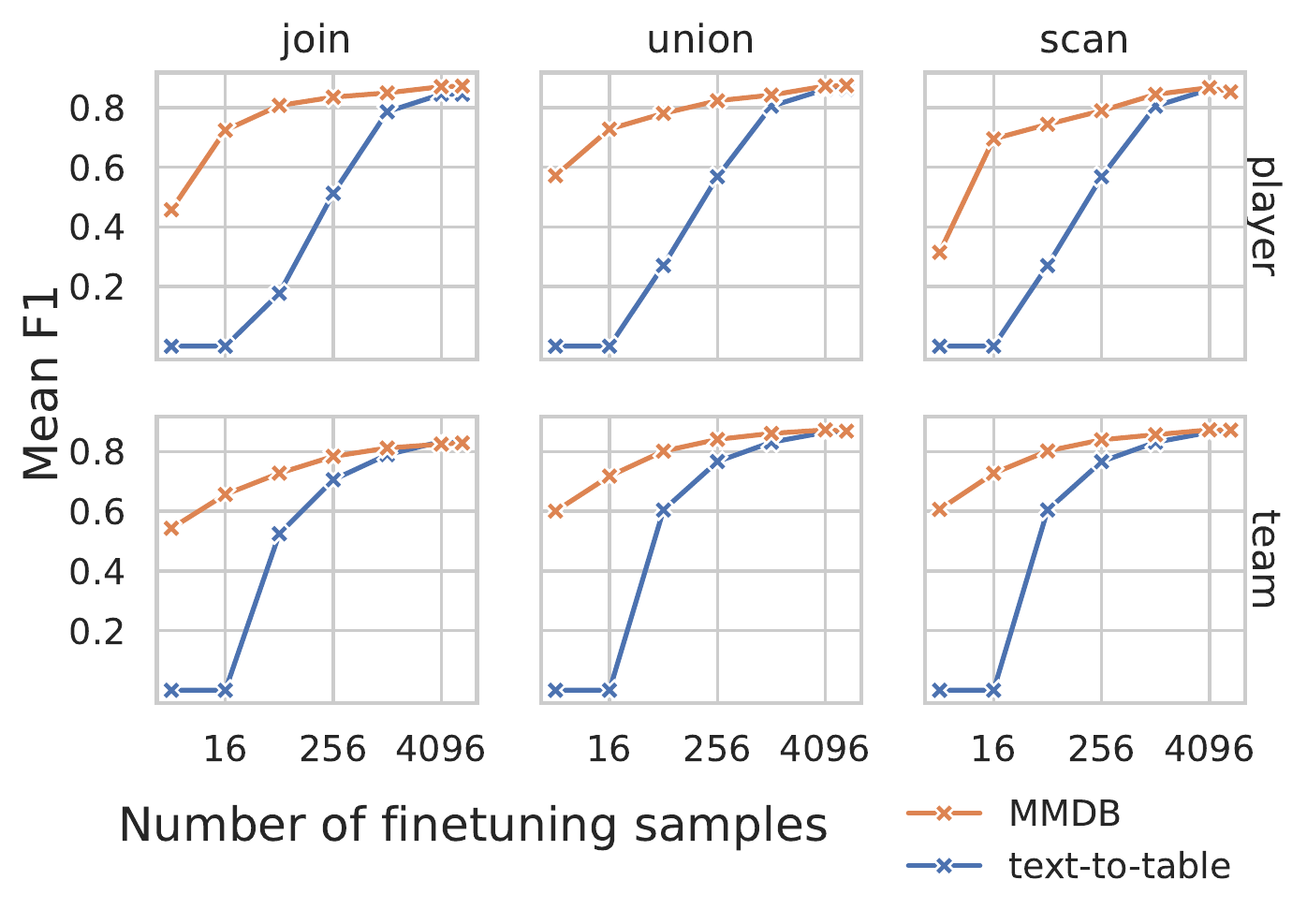}
    \vspace{-4.5ex}
    \caption{Mean F1 scores for MMOps on the \emph{rotowire} dataset. The F1 scores for all models increase as the number of training samples increases, but MMDB consistently outperforms text-to-table.}
    \vspace{-4.5ex}
    \label{fig:rotowire}
\end{figure}

\noindent\textbf{Results on \emph{rotowire}.} The \emph{rotowire} dataset is very different from our pre-training dataset because the texts are game reports, not Wikipedia abstracts and the tables contain numbers only (e.g. the number of scored points).
Hence, the evaluation on \emph{rotowire} can provide a good estimate of our model's performance in a new domain.
Figure \ref{fig:rotowire} shows how the extraction quality increases when we provide more training data for fine-tuning to the new domain.
Even with as few as four labeled texts per table as training data, our model is already able to achieve a mean F1-score of up to 60\% for both the player table and the team table.

Since text-to-table has been pre-trained on plain-text only, it struggles in these few-shot scenarios.
It needs at least 64 labeled texts as training samples to achieve a mean F1 score larger than 0\%.
As the size of the training dataset is increased, we see that the performance of both models increases, but it takes over 4000 training samples before text-to-table matches the MMDB-model.
When trained on the full dataset, MMDB is still slightly better, especially for joins with the player table, which is still impressive, since we did not tune our hyperparameters for the \emph{rotowire} dataset which was used for text-to-table though.
Furthermore, comparing scans to unions, we see that especially with very limited training data (i.e. four training samples), the example values provided in a union can greatly improve the extracted tables.

\begin{figure}
    \centering 
    \includegraphics[width=0.85\columnwidth]{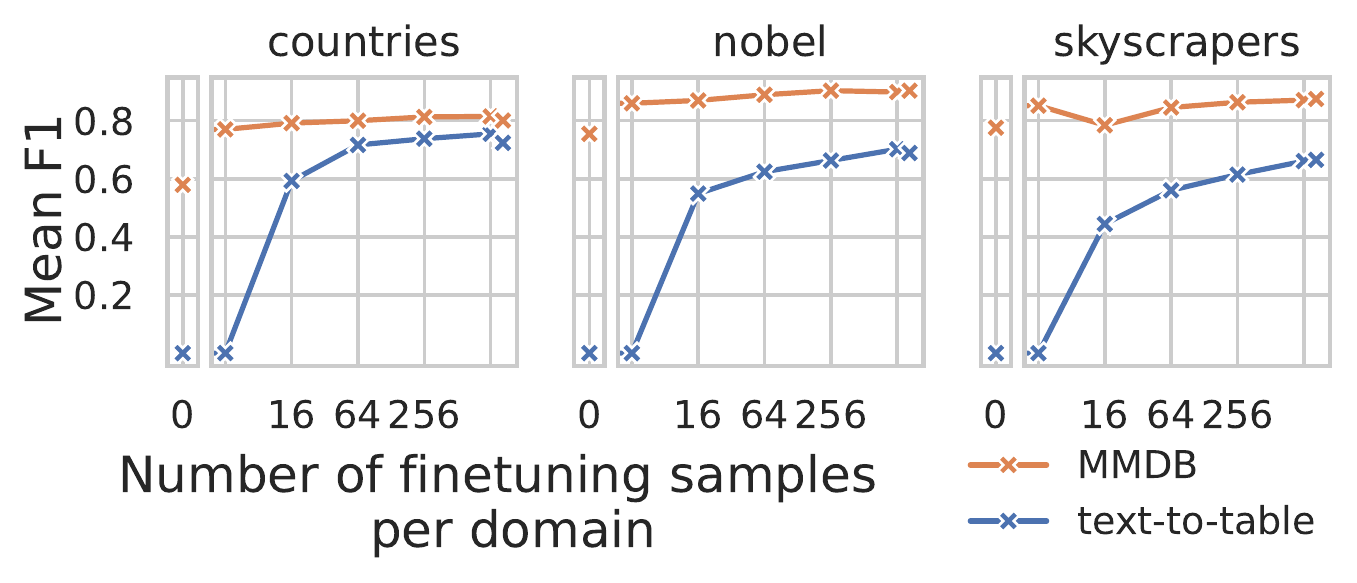}
    \vspace{-4.5ex}
    \caption{Mean F1 scores for a multi-modal scan on different domains of T-REx. The zero-shot performance is depicted on the very left of each plot. MMDB achieves impressive zero-shot performance and outperforms text-to-table even with increasing training data.}
    \vspace{-4.5ex}
    \label{fig:trex}
\end{figure}

\noindent\textbf{Results on T-REx.} The T-REx dataset is more similar to our pre-training dataset since texts are based on Wikipedia abstracts. However, we make sure that it only contains texts not seen during pre-training and thus contains three unseen domains.
Here, the tables contain not only numbers as values but also dates and short spans of texts.
In these scenarios, MMDB can even be used out-of-the-box, in a zero-shot mode without any additional fine-tuning.
Figure \ref{fig:trex} shows how extraction quality improves when increasing the training set size for a multi-modal scan.
With no training data at all, we achieve almost 80\% F1 score for both the nobel and the skyscraper domain.
Moreover, with increasing training data, there is always a large margin between MMDB and text-to-table.
We attribute this to our novel pre-training procedure, which teaches MMDB the necessary skills to perform MMOps.

\subsection{Exp. 2: Runtime of Multi-modal Operators}

The second experiment focuses on the runtime of multi-modal queries in MMDB.
While it is certainly computationally expensive to incorporate a pre-trained language model during query execution, we think there are many scenarios (e.g. queries with a low selectivity) where it makes sense to use a large language model for online (ad-hoc) queries on text collections.
Moreover, MMDB offers extensions such as secondary indexes that make it very effective in these scenarios, as we show in the upcoming experiments.
For all other scenarios, we offer materialized views to remove the computational burden of text processing from query time.
In this experiment, we perform all experiments on the test set of \emph{rotowire}, because it allows measuring the more interesting complex operators.

\begin{figure}
    \centering 
    \includegraphics[width=0.75\columnwidth]{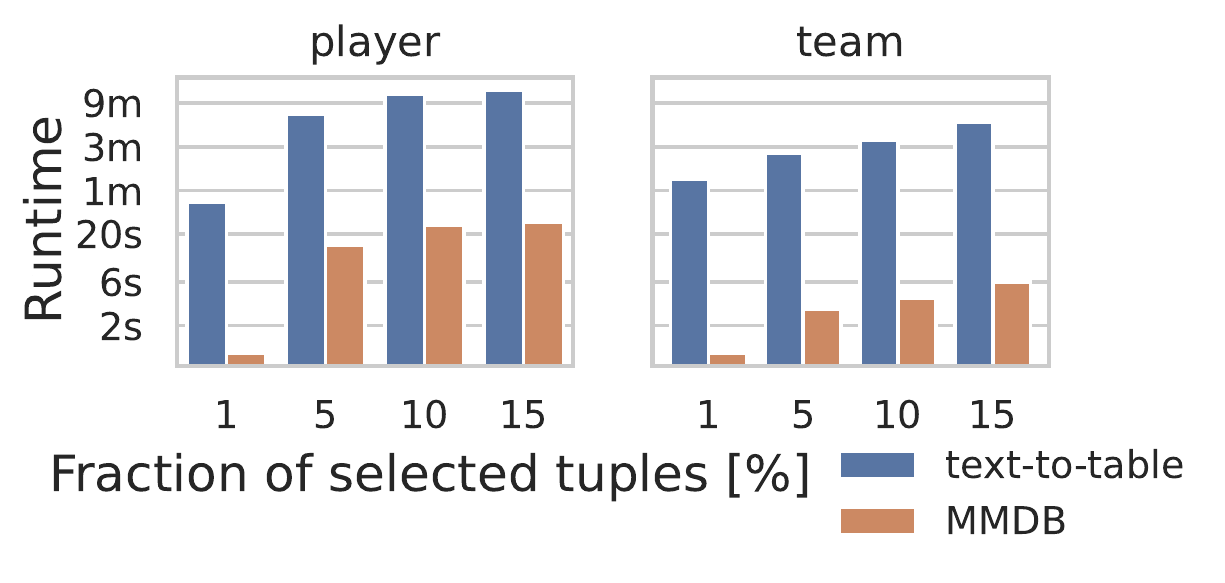}
    \vspace{-4.5ex}
    \caption{Runtimes on \emph{rotowire}. We vary the selectivity of join queries and observe the runtime of the multi-modal operator when implemented with our model or text-to-table. The Y-axis is in log-scale to be able to show the longer execution times of text-to-table (in the order of several minutes) and MMDB (in the order of a few seconds) in the same plot.}
    \vspace{-4.5ex}
    \label{fig:selectivity}
\end{figure}

\noindent\textbf{Selective Multi-modal Join Queries.} In the first scenario, we look at selective multi-modal queries resulting from joins between a table and a text collection when using a filter attribute on the table attributes.
As a dataset, we use the game reports of \emph{rotowire} and join them with the general information tables for \texttt{players} and \texttt{teams} as explained above.
The join queries in this experiment have different selectivities; i.e., the query only selects a few \texttt{players} or \texttt{teams} before executing the multi-modal join with the text collection.
Since tuples in the tables are linked to the game reports (e.g. a tuple about a team is linked to all the game reports the team participated in), this usually results in only a few selected texts as well (i.e., the filter on the table acts as an index into the text collection).
Note, since \texttt{players} are from different \texttt{teams} and participate in many games, for queries that involve the \texttt{players} table, the runtime is overall higher since the join is less selective (i.e., with the same selectivity on the \texttt{players} table, more game reports are selected).

Figure \ref{fig:selectivity} shows how the different selectivities affect the overall query runtime of the query containing the multi-modal join operator.
We compare a version of the query which is implemented naively using a text-to-table model and our MMDB-Model.
The main difference between these approaches is that text-to-table always needs to first translate each selected text into full player and team statistics tables, containing the statistics of all mentioned teams and players.
Since many generated rows will not have a join partner in joins with low selectivity, this leads to many generated rows being discarded.
Our MMDB, instead, is able to generate only those rows that have a join partner using our multi-modal join.
This results in significant differences in the observed runtimes as shown in Figure \ref{fig:selectivity}.
While the operation of text-to-table is usually in the order of several minutes, the multi-modal join is typically able to perform the operation in just a couple of seconds while having higher accuracy as shown in the previous experiment.
For instance, for a join with the \texttt{players} table, text-to-table requires about eleven minutes to perform the operation, while our approach only needs 25 seconds.

\begin{figure}
    \centering 
    \includegraphics[width=0.95\columnwidth]{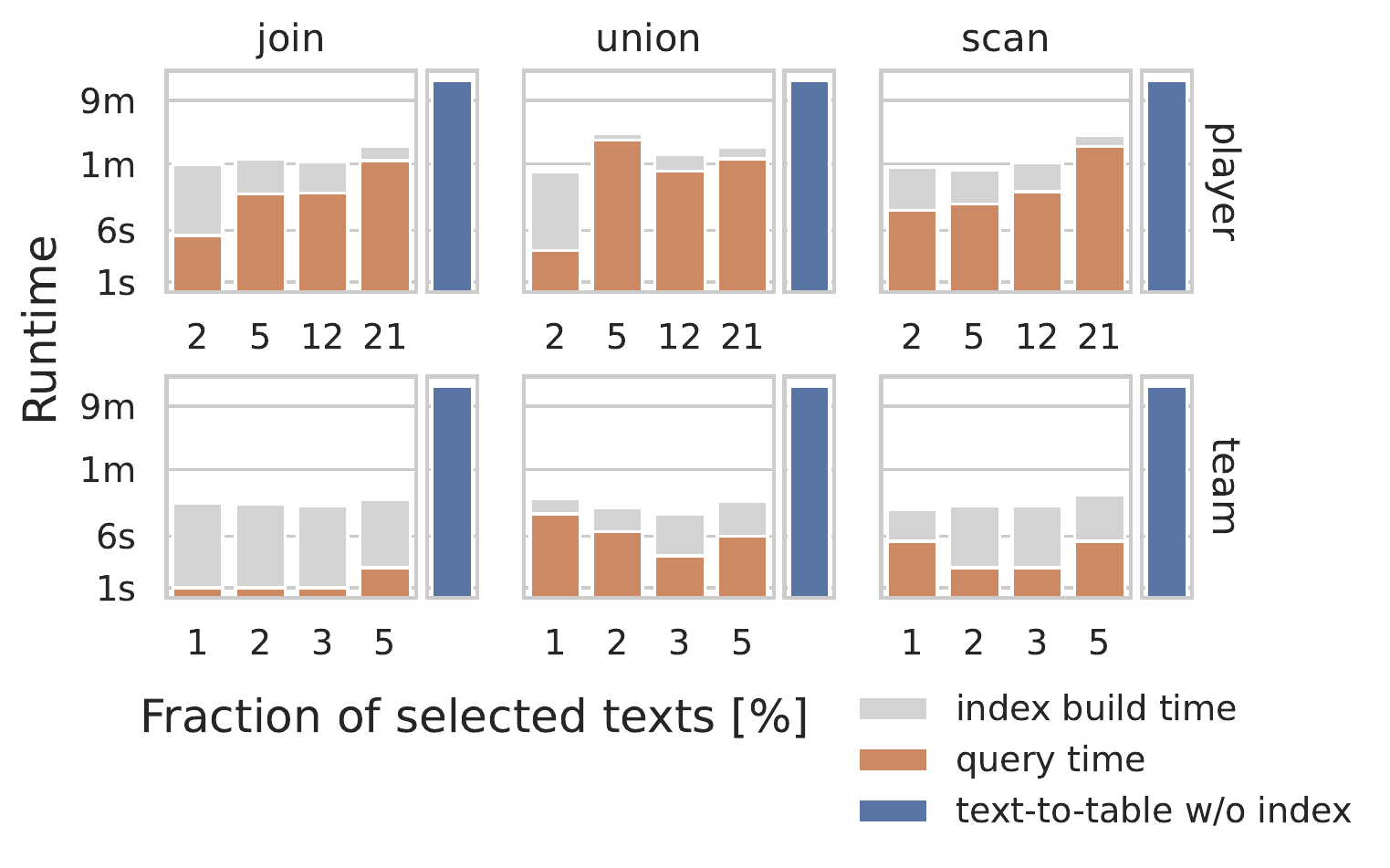}
    \vspace{-4.5ex}
    \caption{Execution times of different MMOps with a multi-modal secondary index. The index allows to only transform those texts to tables, which are selected due to a selection criterion based on a value extracted from the text collection. The Y-axis is in log-scale to be able to plot the naive solution using text-to-table and the runtime when using the index in the same plot.}
    \vspace{-4.5ex}
    \label{fig:index}
\end{figure}

\noindent\textbf{Selective Multi-modal Queries using an Index.} In the second scenario, we analyze the runtime of simple queries that only need to scan a subset of a text collection using a filter operation on one of the extracted attributes.
However, if implemented naively, this query results in a costly operation since independent of the selectivity all texts need to be first transformed to tuples in order to judge which texts qualify for the filter condition.
As a dataset, we again use \emph{rotowire}.
For the queries, we execute different types of queries that can make use of an index on text attributes as introduced in Section \ref{sec:performance-optimizations}: (1) queries with only multi-modal scan, (2) queries that join a general information table with a text collection of game reports, (3) queries which union a statistics table with the game reports. All these queries have a filter with different selectivity on an attribute extracted from the text collection (e.g. number of scored points equals a specified value).

Figure \ref{fig:index} shows the runtime with different filter conditions on differed MMOps, resulting in different amounts of selected texts.
We see that for queries with low selectivity, MMDB is again able to compute the query results in the order of a couple of seconds.
The naive solution of using text-to-table to translate all texts into tables first and then doing the selection afterwards takes much longer.
Moreover, in addition to the query runtime, we show the index creation time which shows that the construction of our index is more efficient than a full scan since it only extracts the values for one column from the texts and does not require the decoding of the full text into a table.
Hence, even if we include the time of building the index, which is usually done offline, we are still an order of magnitude faster than the naive implementation.

\begin{figure}
    \centering 
    \includegraphics[width=0.8\columnwidth]{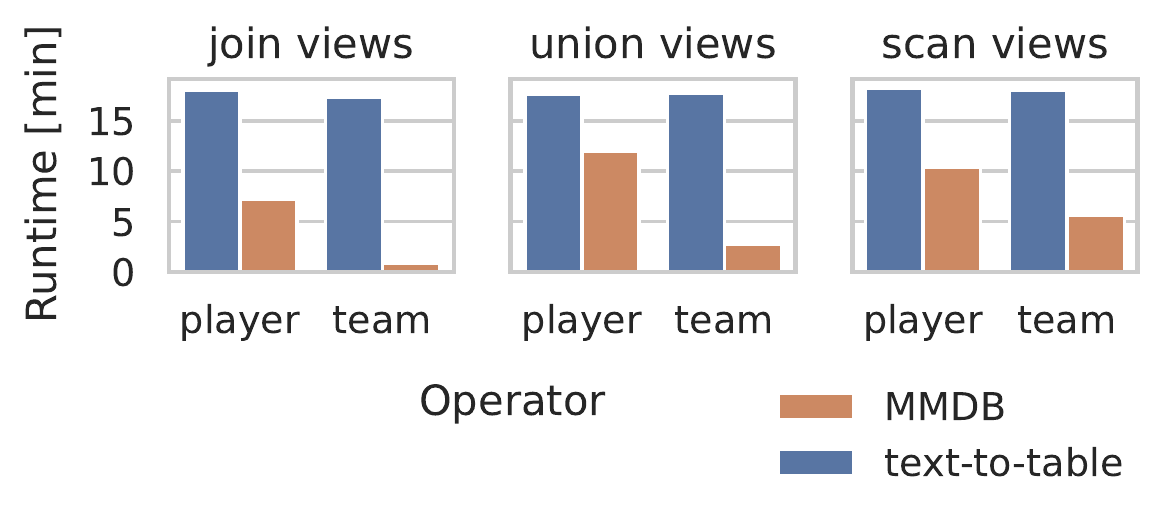}
    \vspace{-4.5ex}
    \caption{Time to construct a materialized view on \emph{rotowire} using MMDB and text-to-table. Text-to-table uses a generative decoder, and thus always needs above 15 minutes for materialized view creation.}
    \vspace{-4.5ex}
    \label{fig:mv}
\end{figure}

\noindent\textbf{Materialized Views.}
While selective queries can be executed in MMDB efficiently as shown before, queries that need to read over large fractions of a text collection will still not be able to be executed efficiently.
For this reason, we introduce multi-modal materialized views that allow doing the heavy lifting of text processing in an offline phase.
A crucial aspect, however, is still the efforts needed to construct materialized views.
In this experiment, we thus compare the cost of constructing materialized views using the queries that are based on our MMDB-Model with an approach that uses text-to-table for the view creation.

For the experiment, we construct a total of six materialized views.
For the first two materialized views, we join the entire text collection of \texttt{rotowire} game reports with the general information tables of the \texttt{teams} (1st view) and \texttt{players} 
 (2nd view).
For the next two views, we union the game reports with the statistics tables for \texttt{players} and \texttt{teams} from the \emph{rotowire} training set.
For the last two views, we transform all game reports to additional statistics tables about players and teams extracted from the texts using multi-modal scans.

As we see in Figure \ref{fig:mv}, using MMDB is much more efficient in creating a materialized view compared to text-to-table.
The main reason is that text-to-table uses a generative decoder, which decodes texts into tables by outputting one token at a time in an autoregressive manner.
Due to the use of self-attention in the decoder, this process is quadratic in the number of output tokens.
Our decoder, however, is based on shallow neural networks, which are able to decode entire rows in one model pass.
Hence, the number of passes through the model is determined by the number of rows to be decoded, which is usually much lower than the number of tokens necessary to describe a table.
As shown in Figure \ref{fig:mv}, we can see that MMDB is especially effective for creating a view for the full \texttt{team} table from the game reports of the \emph{rotowire} dataset.While text-to-table needs over 15 minutes to construct any materialized view on the dataset, MMDB can easily construct such a view in under a minute.

\subsection{Exp. 3: Ablation Study for Pre-Training}\label{sec:ablation}

\begin{figure}
    \centering 
    \includegraphics[width=0.85\columnwidth]{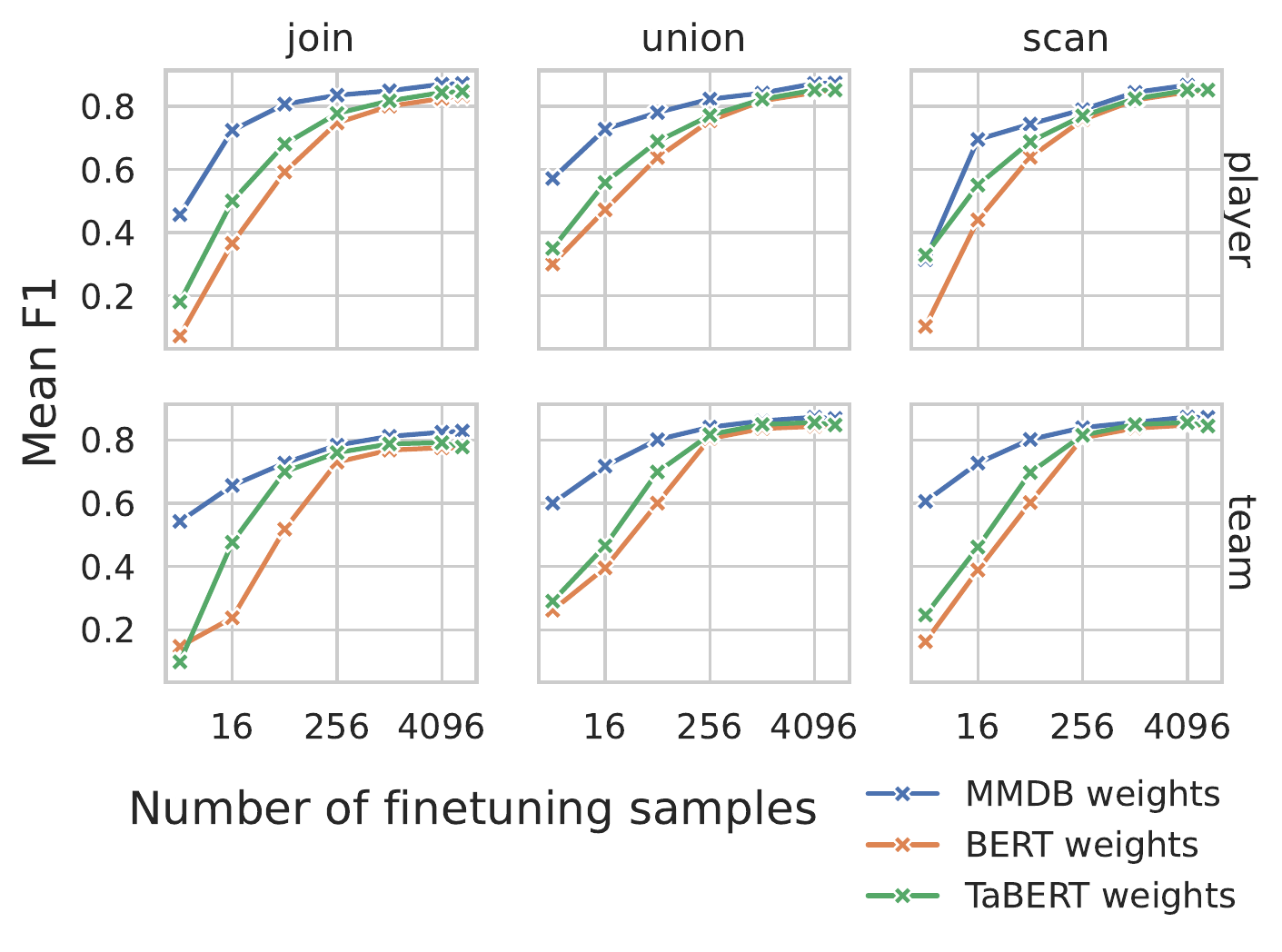}
    \vspace{-4.5ex}
    \caption{Comparison of the query results quality when using MMDB with different pre-trained weights. The pre-trained weights resulting from our pre-training procedure result in better extractions, especially when only limited fine-tuning data is available.}
    \vspace{-4.5ex}
    \label{fig:ablation}
\end{figure}

In the final set of experiments, we show the effect of our pre-training objectives by comparing it to other existing pre-training procedures and by individually disabling our pre-training objectives.

\noindent\textbf{Comparison with other Pre-trained Models.}
The pre-training procedure proposed in this paper aims at teaching the model the necessary skills it needs for performing MMOps.
To examine if this pre-training procedure results in better extractions compared to more traditional pre-training procedures, we compare the quality of query results when using MMDB initialized with different pre-trained weights.
We compare against two pre-trained procedures: the pre-training procedure of BERT \cite{bert} and the pre-training procedure of TaBERT \cite{tabert}.
The BERT model aims at natural language understanding and is thus pre-trained on plain-text only, meaning it has never seen any tabular data before.
Different from BERT, TaBERT's pre-training aims to improve the performance of BERT on tasks like text-to-SQL, where also understanding of tabular data is necessary.
Similar to MMDB, TaBERT is thus also pre-trained on a parallel corpus of texts and tables scraped from the web, and thus already has seen lots of tables during pre-training.

However, in the parallel corpus of TaBERT, texts and tables are often only vaguely related.
Moreover, the pre-training objectives are not designed for table extraction from text and thus are not ideal for enabling MMOps.
Figure \ref{fig:ablation} shows how the different pre-training objectives and corpora used, affect the quality of query results when varying the number of samples for fine-tuning.
Especially in zero-shot or few-shot scenarios with only limited fine-tuning data, our model is consistently able to extract tabular data from text more accurately.
However, surprisingly even with the full \emph{rotowire} training set, there is still an advantage of up to 2.5\% when using our pre-training procedure compared to BERT and TaBERT.
Comparing TaBERT to BERT, we see that TaBERT is able to consistently outperform BERT, which is due to the fact that it has already seen tabular data during pre-training.

\textbf{Effect of Pre-training Objectives.}
In this experiment, we analyze the efficiency of each individual pre-training objective of our MMDB-Model.
Our pre-training procedure consists of three main pre-training objectives, each targeting different skills for table extraction from text.
To show the efficiency of the pre-training objectives, we first disable the \emph{Column-Text-Alignment (CTA)} pre-training objective.
To still be able to perform all database operations with CTA disabled, we use the span-detect head in all cases where the column-detect head has previously been used (i.e. in the first phase of the algorithm to compute complex operators).
Table \ref{table:no-cta-dd} shows the consistently positive effect of enabling CTA.
The reason for this is that the span-detect head, as it is pre-trained using the \emph{Masked Cell Reconstruction (MCR)} objective, tends to only extract values for single entities from texts.
This is a problem for complex texts where multiple rows need to be extracted, since in the first phase, values for multiple entities need to be extracted.

Next, we disable the \emph{Duplicate-Detect (DD)} objective and use simple string matching to detect duplicates when it is disabled instead.
The results are shown in Table \ref{fig:ablation} as well.
Here we see a significant improvement in the countries dataset, where texts often contain the same name of a country, but in different surface forms (e.g. USA and United States).
However, for the other two domains, duplicates that differ in the surface forms seem to be rare and string matching seems to be good enough for detecting them.

We do not disable the MCR objective, because it is impossible compute database operators without the span-detect head.

\begin{table}
\small
\begin{center}
\begin{tabular}{ |c|c|c| }
  \hline
  \textbf{dataset} & \textbf{enabling CTA} & \textbf{enabling DD} \\
  \hline
  \textbf{nobel} & +4.8\% & no improvement \\ 
  \textbf{skyscrapers} & +8.2\% & no improvement \\  
  \textbf{countries} & +6.9\% & +13\% \\
  \hline
\end{tabular}
\caption{Zero-shot F1 improvement of enabling the Column-Text-Alignment (CTA) and Duplicate Detection (DD) pre-training objectives on all three domains of T-REx.}
\label{table:no-cta-dd}
\vspace*{-7.5ex}
\end{center}
\end{table}
 \section{Related Work}\label{sec:related-work}

In the following, we discuss different lines of related work.

\noindent\textbf{Multi-Modal Data Systems.}
There are only a few systems that allow database operations over text or other additional modalities.

A first system, OpineDB \cite{subjective-databases} links subjective texts (e.g. user reviews) to relational data and extends SQL by natural language conditions to select relational data based on texts.
However, OpineDB differs from our work in that texts are only used for selection, but it does not expose text data as tables for query processing as we do.

Another system is NeuralDB \cite{neuraldb}. NeuralDB is a database that uses pre-trained language models to run natural language queries directly on text documents.
Different from MMDB, NeuralDB works only on very short textual statements of a few words.
Moreover, NeuralDB does not consider the  multi-modal case where tabular data is available in addition to texts.

Furthermore, another system is WannaDB \cite{wannadb}, which similar to NeuralDB allows the execution of SQL queries over text collections.
For extracting structured data from text, it uses an interactive matching process based on human feedback.
Unlike our work, however, it again is not multi-modal and has further restrictions such as it assumes that each text document adds only a single row to the output table (i.e. no support for Multi-Row-Texts) and it also does not support more complex operations such as joins and unions of tables and texts.

Recently, \citet{symphony} presented their vision of multi-modal datalakes, Symphony, which can be queried using natural language queries.
The setting in multi-modal datalakes is different from databases, since the main concern is retrieving data from multi-modal datasets.
For retrieval, they propose a self-supervised information compression pre-training objective to embed multiple modalities in the same latent space and retrieve based on similarity in this latent space.
After retrieval, they use different approaches to answer the user query depending on the modality.
For text data, their model is based on NeuralDB.

\vspace{0.5ex}\noindent\textbf{Pre-training Models.} Large pre-trained language models \cite{elmo, bert, roberta, gpt3} are by now dominating NLP and are quickly adapted for multi-modal \cite{vilbert, vl-bert, lxmert, visual-bert} and tabular data \cite{turl, tabbie, tuta}.
To alleviate low overhead adaption to downstream tasks, pre-training objectives began to be more aligned with the downstream task for many core-NLP \cite{sspt, marge, span-bert-a, span-bert-b} and also structure-aware tasks \cite{tapex, grappa, gap, mqa-qg}.

Most similar to our pre-training objectives are those pre-training procedures that rely on weak or distant supervision to align pre-training more to the downstream task.
ReasonBERT \cite{reason-bert} uses a pre-training objective inspired by distant supervision for the downstream task of multi-hop hybrid question answering.
StruG's \cite{strug} pre-training dataset designed for the text-to-SQL task is based on the table-to-text generation dataset ToTTo \cite{totto} that was extracted from Wikipedia using heuristics.
ToTTo contains 120.000 training samples and is thus much smaller than our pre-training dataset.

\vspace{0.5ex}\noindent\textbf{Extraction of Tabular Data from Text.}
Text-to-table \cite{text-to-table} is a sequence-to-sequence model trained to transform tables into text.
It introduces several model adjustments to ensure that the model outputs a correctly structured table.
\citet{stable} argue that in some cases decoding from left to right and top to bottom is not optimal and thus propose STable, a model similar to text-to-table, which is able to output table cells in arbitrary order.
Unlike our work, text-to-table and STable are trained in a supervised manner from scratch for every new dataset.

\vspace{0.5ex}\noindent\textbf{Cross-Domain Slot Filling.}
Slot filling is used in task-oriented dialog systems to fill a set of slots (e.g. depart, arrive) from a natural language utterance (i.e., texts).
This is similar to basic database operations which extract rows from text, where values for a set of column names have to be extracted from text.
Some recent works also use pre-training for slot filling while mostly relying on conversational data from Reddit \cite{convex, gensf}.
Different from our setting, the texts tend to be rather short and limited in variability.
Moreover, again complex multi-modal operations such as joins and or unions are not supported by these approaches.

 \section{Conclusions and Future Work}
\label{sec:conclusion}

In this paper, we presented  multi-modal databases, which is a new class of databases that allow users to seamlessly query textual and tabular data.
The cornerstone of multi-modal databases are multi-modal database operators, which we have shown to be realizable using large pre-trained language models.
As a result, multi-modal database operators can be executed on new datasets from unseen domains in a zero-shot or few-shot manner with no or only minimal fine-tuning overhead.
In the future, we also want to explore how other modalities like  images could be incorporated in our database to enable database operations for even more modalities.

\begin{acks}
We thank the reviewers for their feedback. This research is funded by the Hochtief project \emph{AICO} (AI in Construction), by the BMBF and the state of Hesse as part of the NHR Program, as well as the HMWK cluster project \emph{3AI} (The Third Wave of AI). Finally, we want to thank hessian.AI at TU Darmstadt as well as DFKI Darmstadt.
\end{acks}

\clearpage

\balance{}
\bibliographystyle{ACM-Reference-Format}
\bibliography{bib}

\end{document}